\documentclass[
reprint,prb,aip
]{revtex4-1}
\usepackage{caption}
\usepackage{graphicx}
\usepackage{dcolumn}
\usepackage{amsmath}
\usepackage{bm}
\usepackage{tensor}
\usepackage{textcomp}
\usepackage[T1]{fontenc}

\begin{document}


\title{Spatial Sampling of Terahertz Fields with Sub-wavelength Accuracy via Probe Beam Encoding}

\author{Jiapeng Zhao}
\affiliation{The Institute of Optics, University of Rochester, Rochester, New York, 14627, USA}
\author{Yiwen E}
\affiliation{The Institute of Optics, University of Rochester, Rochester, New York, 14627, USA}
\author{Kaia Williams}
\affiliation{The Institute of Optics, University of Rochester, Rochester, New York, 14627, USA}
\author{Xi-Cheng Zhang}
\email{xi-cheng.zhang@rochester.edu}
\affiliation{The Institute of Optics, University of Rochester, Rochester, New York, 14627, USA}
\author{Robert W. Boyd}
\email{boydrw@mac.com}
\affiliation{The Institute of Optics, University of Rochester, Rochester, New York, 14627, USA}
\affiliation{Department of Physics, University of Ottawa, Ottawa, K1N 6N5, Canada}

\date{\today}
\maketitle
\begin{quotation}



 Recently, computational sampling methods have been implemented to spatially characterize terahertz (THz) fields. Previous methods usually rely on either specialized THz devices such as THz spatial light modulators \cite{chan2009spatial,watts2014terahertz}, or complicated systems requiring assistance from photon-excited free-carriers with high-speed synchronization among multiple optical beams \cite{ulbricht2011carrier,shrekenhamer2013terahertz, stantchev2016noninvasive}. Here, by spatially encoding an 800 nm near-infrared (NIR) probe beam through the use of an optical SLM, we demonstrate a simple sampling approach that can probe THz fields with a single-pixel camera. This design does not require any dedicated THz devices, semiconductors or nanofilms to modulate THz fields. Through the use of computational algorithms, we successfully measure 128$\times$128 field distributions with a 62 $\mu m$ transverse spatial resolution, more than 15 times smaller than the central wavelength of the THz signal (940 $\mu m$). Benefitting from the non-invasive nature of THz radiation and sub-wavelength resolution of our system, this simple approach can be used in applications such as biomedical sensing, inspection of flaws. in industrial products, and so on \cite{pickwell2006biomedical,karpowicz2005compact}.
 \par
\end{quotation}


\section{Introduction}
The unique properties of terahertz (THz) radiation, such as the high transmittance in nonpolar material and the nonionizing photon energies, enable numerous novel possibilities in both fundamental research and industrial applications
\cite{zhang2010introduction, lee2009principles,karpowicz2005compact,wang2004metal, pickwell2006biomedical, kawase2003non, eisele2014ultrafast}. Consequently, the knowledge of the spatial profile of THz fields becomes very important. However, due to the lack of efficient and economical THz cameras \cite{escorcia2016uncooled,grossman2010passive}, characterizing the structures of THz fields usually relies on raster scanning either the detector or the sample, resulting in a low signal-to-noise ratio (SNR) and slow speed when the number of pixels increases. Recently, novel beam profiling approaches that involve computational sampling methods have emerged \cite{watts2014terahertz,chan2009spatial, ulbricht2011carrier, shrekenhamer2013terahertz}. Computational sampling methods, combining the computational algorithms with optical imaging techniques, can improve the sampling speed and image quality, particularly for weak illumination. These computational methods usually do not require conventional cumbersome techniques such as multi-element THz detection arrays and raster-scanning, but reconstruct the fields by computational algorithms with a single-pixel camera. Nevertheless, much of previous work has been based on spatially manipulating the THz beam directly with either dedicated THz spatial light modulators (SLMs) \cite{chan2009spatial,watts2014terahertz}, or modulating the THz spatial transmittance in semiconductors or nanofilms via photon-excited free carriers \cite{ulbricht2011carrier,shrekenhamer2013terahertz,stantchev2016noninvasive, chen2019terahertz}. Compared to optical SLMs, their THz counterparts usually have a low temporal modulation rate and a large pixel size, which is mainly limited by the relatively long THz wavelengths and leads to a relatively low spatial modulation accuracy and resolution \cite{chan2009spatial,watts2014terahertz}. Therefore, THz SLMs are not ideal devices for sampling applications that requires sub-wavelength features. Although approaches that use spatial transmission modulation via photon-excited free carriers can provide precise spatial modulations up to few microns, these methods require significant laser powers in order to excite free-carriers, as well as high-speed synchronization among multiple optical beams \cite{stantchev2016noninvasive}. Thus, it is important to develop a simpler THz spatial sampling approach capable with up to a micron-level accuracy for practical application scenarios. \par

\begin{figure*}[ht!]
\includegraphics[width=0.9\textwidth,keepaspectratio]{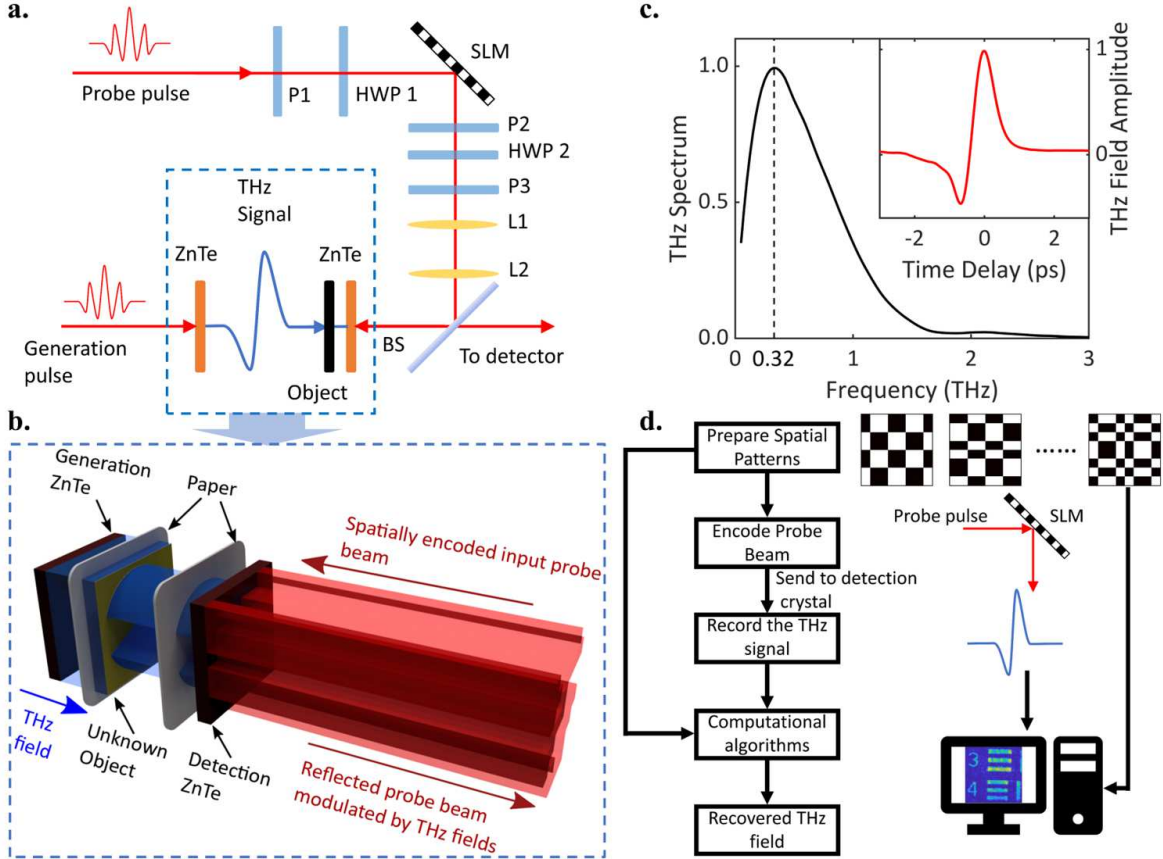}
\caption{(a). Diagram showing the schematic system configuration while (b) shows the detailed feature of our sampling approach. The unknown object, actually an AF test chart, is wrapped in paper so that it is invisible to the NIR probe beam. P1, P2, and P3 are polarizers while HWP1 and HWP2 represent half-wave plates. L1 and L2 image the SLM plane onto the left surface of detection ZnTe crystal. BS is a 50/50 beam splitter. The modulated probe beam is then sent to a single-pixel detector for measurement. Detailed specifications can be found in the Methods. (c). The spectrum of the generated THz pulse, and the time-domain electric field distribution. The spectral peak occurs at 0.32 THz, which corresponds to 940 $\mu m$ central wavelength. (d). The flowchart shows our procedure of single-pixel imaging via probe beam encoding.}
\end{figure*}

The electro-optic (EO) effect is one of the most widely used THz techniques for coherent detection \cite{wu1995free}. When a THz field interacts with an EO crystal (usually ZnTe or GaP), it introduces a birefringence, which modifies the polarization of a co-propagating 800 nm NIR probe beam. This rotation in polarization is measured to determine the time-dependent THz electric field. Therefore, only THz fields that spatiotemporally overlap with the NIR probe beam result in a measurable polarization shift. From a sampling point of view, when we encode the NIR beam with desired patterns and carefully align these patterns with THz fields, those fields of interest can be selectively measured. Since we only spatially modulate the NIR probe beam, this indirect measurement is not only much easier than manipulating the THz fields directly, but also non-demolished to THz spatial information as well. This non-demolition nature may enable the possibility of THz quantum measurements in the future. Furthermore, considering that all the spatial manipulation comes from comparatively economical and well-developed optical SLMs, achieving fast temporal modulation rates and sampling accuracies at few microns is also feasible \cite{edgar2015simultaneous}. Therefore, real-time sensing through the use of our probe-beam-encoding technique should be achievable in the future.\par

In this paper, we sample the spatial distribution of THz fields by encoding an 800 nm NIR probe beam through the use of a NIR SLM that provides a fast sampling rate up to the kilohertz (kHz) level, and a sampling accuracy up to few microns. As a demonstration, we impress a series of masks onto a NIR probe beam, and pass a THz beam through an object that is opaque to the NIR probe. Through the EO effect, the spatial information of the THz beam is transferred to the NIR probe whose total power is then measured. After repeating this procedure for each spatial mask, the THz field distribution carrying near-field information is successfully retrieved with 128$\times$128 sample points through the use of Hadamard Matrix (HM) algorithm \cite{harwit2012hadamard}. The spatial resolution is estimated to be 62 $\mu m$, which is 15 times smaller than the THz central wavelength (940 $\mu m$ at 0.32 THz). By adopting the compressed sensing (CS) algorithm, we can recover high fidelity fields (near 90$\%$ fidelity) with a 20$\%$ sampling ratio. This simple technique provides up to few micron sampling accuracy and a sub-wavelength resolution while inheriting most advantages of THz sensing, such as broadband spectrum information and non-invasive detection to biomedical samples.\par

\section{Experimental configuration and resolution estimation}

\begin{figure*}[ht!]
\includegraphics[width=0.9\textwidth,keepaspectratio]{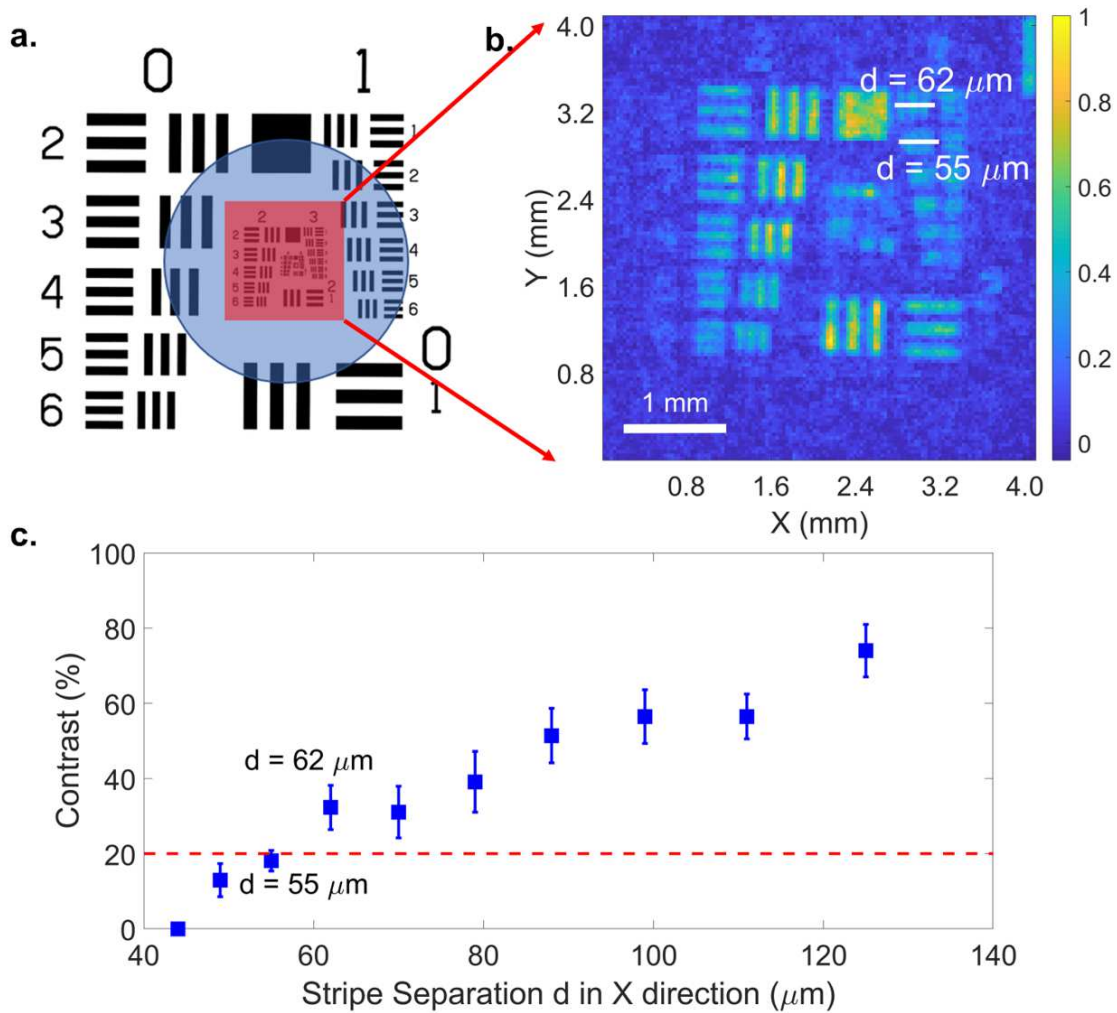}
\caption{(a). The original AF target. The red square is the area that the probe beam illuminates. The blue circle shows the parts illuminated by the THz field. Through the change of the location of the encoding masks on the SLM, we can intentionally probe any part of the THz field in the blue circle without changing the optics. (b). The experimental result with 128$\times$128 pixels showing the resolution limit. (c). The contrast as a function of strip separation d. The red dashed line is the threshold contrast assumed by Rayleigh criterion. Two points can be resolved if the contrast is equal or greater than 20$\%$. Therefore, the element set with d = 55 $\mu m$ is not resolved while the element set with d = 62 $\mu m$ is resolved according to the Rayleigh criterion.}
\end{figure*}

A THz pulse, generated through optical rectification from a ZnTe crystal as shown in the schematic Fig. 1(a) and (b), passes through a covered object and is detected through electro-optic sampling by another ZnTe crystal \cite{xie2006terahertz}. A sequence of spatial masks are loaded on the SLM to encode the NIR probe pulse. Then this spatially encoded 800 nm probe beam first illuminates the detection ZnTe crystal in the counter-propagation direction of the THz pulse as shown in Fig. 1(b). On reflection from the left surface of the detection ZnTe crystal, the probe beam co-propagates with the THz pulse, and its polarization is modulated by this spatiotemporally coincident THz field in the detection ZnTe crystal. This change in the polarization is measured to retrieve the time-dependent THz signal. The object, a positive US Air Force (AF) target made with chromium, is wrapped in a 70-$\mu m$ thick piece of paper, and placed immediately before the detection crystal. Therefore, THz pulse only travels about 70 $\mu m$ before interacting with the detection crystal, and the near-field information is maintained \cite{kowarz1995homogeneous}. The original time domain THz pulse and corresponding spectrum are shown in Fig. 1(b) and (c), respectively. As shown in the flowchart in Fig. 1(d), after retrieving the time-dependent THz signal and recording the corresponding patterns, the THz field information can be reconstructed with computational algorithms.\par

\begin{figure*}[ht!]
\includegraphics[width=0.6\textwidth,keepaspectratio]{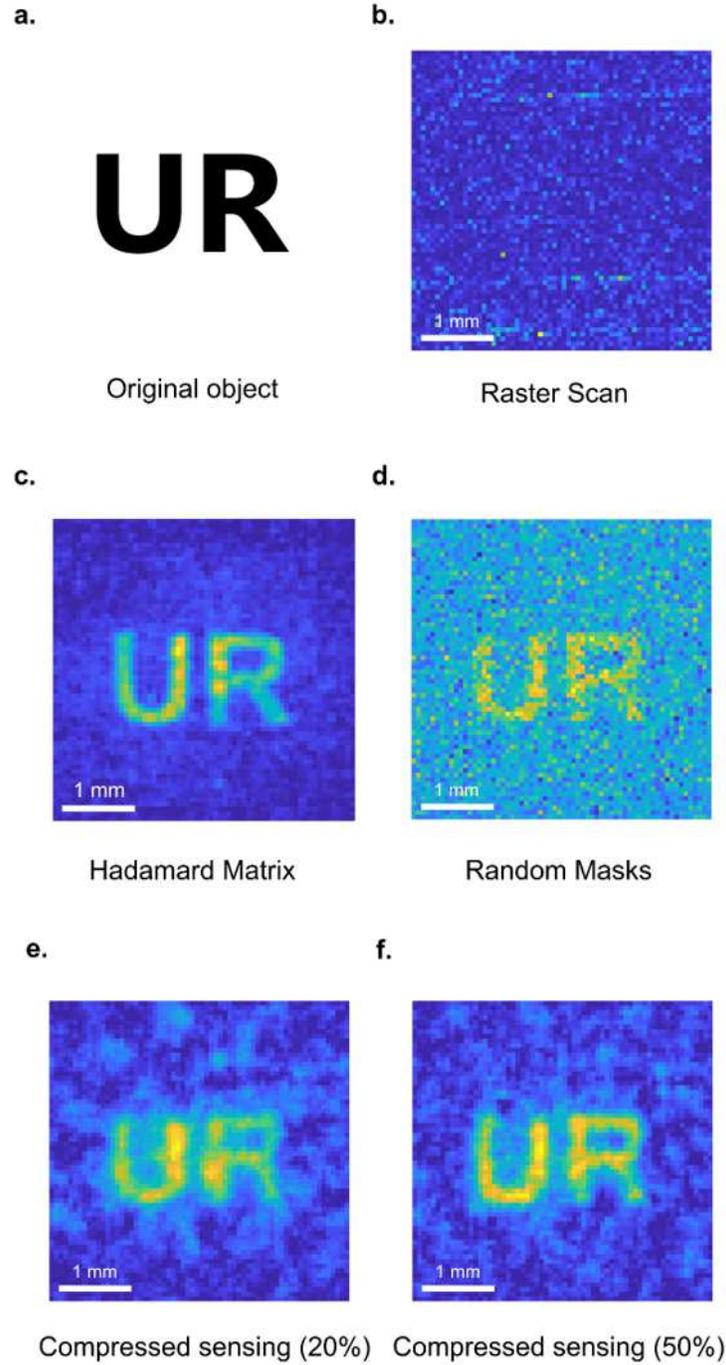}
\caption{(a). Original \textquotedbl UR\textquotedbl\hspace{0.05cm} object. (b). Field distribution obtained by raster scanning. (c). Field distribution reconstructed by HM algorithm. The bright central part of the \textquotedbl UR\textquotedbl\hspace{0.05cm} comes from the non-uniformly distributed THz field. (d). Reconstructed field obtained by random masks. (e) and (f). Recovered field through the use of CS with different sampling ratios. All recovered fields have 64$\times$64 sampling points with 64 $\mu m$ pixel size. }
\end{figure*}

We first use the HM algorithm to probe the THz field, and successfully recover the field distribution with 128$\times$128 sample points. As shown in Fig. 2(a), we selectively sample the central elements of the AF chart to estimate the resolution limit. The recovered intensity is shown in Fig. 2(b), and has a square sampling pixel with 32 $\mu m$ width. In Fig. 2(c), we show the contrast of each element in group 2, and element 3-1 (the first set of elements in group 3) and 3-2 as a function of strip separation d in the X direction. We find that the element with d = 55 $\mu m$ (element 3-2) has a average contrast of 18.10$\%$, while the element with d = 62 $\mu m$ (element 3-1) has a 32.28$\%$ average contrast. Since from Rayleigh criterion, two points are barely resolved with a contrast equal to 20$\%$ (red dashed line in Fig. 2(c)), the resolution of our system is found to be 62 $\mu m$. Considering the central THz wavelength ($\lambda_c$) is 940 $\mu m$, we have achieved a resolution of about $\lambda_c /15$. It should be noted that the features in the X direction are better resolved than those in the Y direction due to the horizontal polarization of the THz beam \cite{stantchev2016noninvasive}. A better resolution in the Y direction can be expected when the THz field is vertically polarized. Since the strip separations in elements 3-1 and 3-2 are 62 $\mu m$ and 55 $\mu m$ but the sampling pixel is 32 $\mu m$, we can see strong pixelization effects in the experimental data. The pixelization makes the reconstructed image blurred, and further limits the resolution of our scheme. Through the simulations shown in the Supplementary Material, we find that the resolution of our configuration (with a 32 $\mu m$ pixel size) is less than 35 $\mu m$. Another interesting observation from the simulation shows that a longer central wavelength can provide a better spatial resolution. This counter-intuitive conclusion comes from the nature of near-field imaging. As analyzed in Ref. \cite{kowarz1995homogeneous}, for the diffraction field of a sub-wavelength object in the near-field region, a small ratio of $z/\lambda$ can maintain more features, where $z$ is the propagation distance. Therefore, the factors that limit the spatial resolution include the pixel size, the thickness of the detection crystal, the central wavelength of the THz pulse and the separation between the detection crystal and object. Thus, we can further improve the resolution to few microns through illuminating a longer wavelength THz pulse on a thinner crystal, as well as encoding the probe beam with a smaller pixel size and moving the sample closer to the detection crystal \cite{mitrofanov2006tetrahertz}.

Note that, our scheme can probe any portion of the THz field without changing optics. This is because that we can change the location of the NIR probe on the THz field through the change of the location of the encoding masks on the SLM using Matlab. For instance, the results that we show above use the central part of the SLM for encoding, which yields a reconstructed field of the central part of THz field. \par

\section{Results with different computational algorithms}

In the traditional THz beam profiling, raster scanning is the prevalent single-pixel sampling technique due to the lack of economical high performance cameras. The limitations in speed and contrast become apparent when the total number of sampling points increases. With a finer sampling and an increased number of pixels, the SNR on each pixel is reduced due to the reduction in the signal level on each pixel. As a result, one needs to significantly increase the integration time in order to average out the noise. Furthermore, finer sampling also requires very precise mechanical controls. To overcome these limitations, computational techniques that project multi-pixel spatial masks including random patterns and HM are introduced, which remove the requirement of mechanical scanning and result in an accurate sampling control. Because multiple pixels are sampled in each measurement, the limitations due to the detector noise are mitigated. Moreover, different algorithms can provide various benefits in image quality and reconstruction speed. For example, when the source noise is negligible, the HM algorithm that minimizes the mean squared error gives the best SNR \cite{harwit2012hadamard}, and CS can reconstruct the field with sub-Nyquist sampling rates for a fast measurement (see Supplementary Material) \cite{li2011compressive}. \par

The experimental comparison of different strategies is shown in Fig.3. The object is a positive \textquotesingle UR\textquotesingle \hspace{0.05cm} mask (Fig. 3(a)) with a 300 $\mu m$ line width ($\lambda_c/3.13$). In comparing the recovered normalized field distributions obtained from raster scanning (Fig. 3(b)), random binary masks (Fig. 3(d)), HM (Fig. 3(c)) and CS (Fig. 3(e) and (f)), we see that all the computational algorithms have a better SNR than raster scanning for the same acquisition time. This is because for the pixel size of 64 $\mu m$, which is much smaller than the wavelength, the signal of the THz field on each single pixel is less than the detector noise. Therefore, the results measured by raster scanning only gives the detector noise but not reveal any spatial information about the THz field. Comparing the results from all computational algorithms, HM provides the best contrast, while the image reconstructed from random masks is the noisiest, which matches the expectation. By sub-sampling the field with a $20\%$ sampling ratio, the CS yields a field distribution with 89.6$\%$ fidelity. Fidelity is defined as the correlation coefficients between the recovered THz field and the original object. With a 50$\%$ sampling ratio, we can achieve $96.9\%$ fidelity, which is mainly limited by the background noise. One can further improve the fidelity by adding more image processing algorithms in the restoration stage, which is beyond the scope of this paper. Therefore, high fidelity field distributions are accessible with less than half of the original total measurement time by combining CS algorithms with more data processing algorithms. A faster acquisition time is mainly limited by the low switching speed of the SLM (60 Hz), and will be further limited by the repetition rate of our laser (1 kHz) if a high-speed (kHz-level) digital micromirror devices (DMDs) is used as the new SLM. Since a THz field with a uniform spatial distribution is the only requirement for field reconstruction with high quality, spintronic THz emitters pumped by a high repetition rate oscillator laser can be employed for ultrafast sampling \cite{seifert2016efficient}, which can possibly leads to real-time beam profiling \cite{edgar2015simultaneous}.

\section{Discussion and Conclusion}
We have demonstrated that our near-field spatial sampling technique through the use of a spatially encoded probe can have a better sampling accuracy, resolution, and contrast in comparison to raster-scanning methods.
These advantages are facilitated by computational algorithms that offer a general advantage over all THz imaging methods based on raster scanning \cite{watts2014terahertz}. Not only the amplitude distribution , but the phase and spectral distributions can also be extracted by recording the time-dependent THz waveforms \cite{chan2008single,stantchev2018subwavelength}. In relation to the EO imaging \cite{jiang19992d}, we do not measure the THz spatial profile directly, but recover the field distribution through the use of computational algorithms. This non-demolished measurement gives rise to a better performance in both resolution and contrast without requiring high power lasers, especially when the detection crystal is thick (see Supplementary Material) \cite{mittleman2018twenty}. Additionally, indirectly sampling the THz field can also circumvent the requirements of using high-energy lasers, which is a common requirement of most THz sub-wavelength imaging techniques. Furthermore, compared to the imaging techniques using photon-excited free carries, the probe beam encoding also waives the reliance on complicated high-speed synchronization among three arms, which makes the system more reliable \cite{stantchev2016noninvasive, zhao2014terahertz}. Therefore, due to the concise and robust configuration, it is possible to integrate our method into a plug-and-play system with high performance but low cost THz spintronic emitters, leading to numerous possible applications.\par

In summary, we demonstrate a simple method to spatially sample THz fields with up to kHz level sampling rates and a sampling accuracy of few microns. With this approach, we demonstrate a THz near-field sampling system using a single-pixel THz detector. The THz field after an object is successfully measured with 62 $\mu m$ ($\lambda_c /15$) resolution. By adopting the CS algorithm, we can recover high fidelity field distributions (near 90$\%$ fidelity) by sub-sampling the THz beam, providing a way to achieve fast beam profiling. Our approach can retrieve the spatial information of THz field without distorting it. Such a tool can be used in lossless beam profiling, biomedical sensing, flaw detection and security inspection.\par


\section*{Acknowledgement}
The project is funded by Army Research Office (ARO) (W911NF-17-1-0428). We acknowledge the helpful discussions with Saumya Choudary, Yiyu Zhou and Qi Jin. Jiapeng Zhao thanks Brian McIntyre and James Mitchell for fabricating the samples. Robert.W.Boyd acknowledges support from Canada Research Chairs Program and National Science and Engineering Research Council of Canada.\par

\section*{Methods}
An 800 nm Ti-sapphire amplifier laser (Coherent Legend Elite Duo with seed laser Coherent Vitara S) with 1 kHz repetition rate is used. The pulse duration is measured to be 100 fs by an autocorrelator. The collimated beam is split by a 90/10 beam splitter (BS1) at the front of the setup. After the delay line, the pump beam illuminates a $10\times10\times1$ mm ZnTe crystal to generate the THz pulse using optical rectification \cite{xie2006terahertz}. A 10-mm-diameter iris is used before the generation crystal to select the center part of the beam, which gives a relatively uniform intensity distribution of the pump beam. The power used to generate THz is about 1.27 $\text{W/}\text{cm}^2$ corresponding to 1 mJ pulse energy. The generated THz beam illuminates the \textquotesingle unknown\textquotesingle \hspace{0.05cm} object which is wrapped in a 70 $\mu m$ thick paper sheet after going through a silicon wafer that blocks the residual 800 nm pump beam. The paper is opaque to 800 nm and visible light. The NIR probe beam goes through a 15-mm-diameter iris, and is imaged onto the SLM screen (Hamamatsu LCOS-SLM X10468-02). The imaging system consists of one 25-cm focal-length lens and one 20-cm focal-length lens (not shown in the figure). Therefore, the beam diameter on the SLM is 12 mm, which is larger than the THz beam diameter. The SLM spatially encodes phase-only patterns onto the probe arm, and we then use a common-path interferometer to transfer these phase-only patterns to intensity patterns. The common-path interferometer consists of two polarizers (P1 and P2) and the HWP1. P1 is located before the SLM to make sure polarization is horizontal. The half-wave plate 1 (HWP1) then rotates the polarization to $45^\circ$. P2 is also set to $45^\circ$ and located after the SLM. Another HWP2 is used after P2 to rotate the polarization back to the horizontal direction. In the experiment, we observe that when we switch the patterns on the SLM, the polarization state of the beam has also slightly changed, which leads to a noise on the balanced detector. Therefore, two Glan-Taylor polarizers (GCL-0702) (P3) are used to increase the extinction ration in the horizontal direction so that all the vertical polarization get rejected after P3. A 25-cm focal-length lens (L1) and a 20-cm focal-length lens (L2) form an imaging system to image the SLM plane onto the left surface of the detection ZnTe crystal. On reflection from a 50/50 beam splitter (BS), the probe beam carrying spatial patterns goes to the detection ZnTe. The translation stage is well aligned to make sure that the probe pattern spatially overlaps with the THz field of interest, and temporally overlaps with the peak position of the THz pulse. The intensity in the probe beam is about 0.1 $\text{mW/}\text{cm}^2$ corresponding to 16.8 nJ pulse energy.\par

The unknown object to identify the resolution limit is a positive US Air Force target which matches MIL-S-150A standard (Thorlabs R3L3S1P). The target is made from 120-nm-thick chrome deposited onto a 1.5-mm-thick clear soda lime glass substrate. The surface with the test strips faces the detection crystal so that the THz field carrying the object information travels only 70 $\mu m$ before interacting with the detection ZnTe crystal, whose dimensions are $10\times10\times0.1$ mm. \par

To get a fair comparison, the acquisition times of all reconstructed fields in the Fig. 3 are set to be same. The details of how we probe the field using different algorithms can be found in the Supplement Material. Here we need to emphasis that the brighter center part of the recovered fields is caused by the non-uniform spatial distribution of the THz field which has a strong beam center.\par

The \textquotesingle UR\textquotesingle \hspace{0.05cm} sample is positively fabricated with 100 nm thick chromium on a 170 $\mu m$ thick coverslip via physical vapor deposition (PVD). The sample is wrapped in the paper with characters facing the detection ZnTe which gives about 70 $\mu m$ separation between sample and the detection crystal. All the experiments are done in room temperature so that this technique should fit in most applicable scenarios directly.\par


\end{document}


\preprint{APS/123-QED}

\title{Supplementary Material}

\author{Jiapeng Zhao}
\affiliation{The Institute of Optics, University of Rochester, Rochester, New York, 14627, USA}
\author{Yiwen E}
\affiliation{The Institute of Optics, University of Rochester, Rochester, New York, 14627, USA}
\author{Kaia Williams}
\affiliation{The Institute of Optics, University of Rochester, Rochester, New York, 14627, USA}
\author{Xi-Cheng Zhang}
\email{xi-cheng.zhang@rochester.edu}
\affiliation{The Institute of Optics, University of Rochester, Rochester, New York, 14627, USA}
\author{Robert W. Boyd}
\email{boydrw@mac.com}
\affiliation{The Institute of Optics, University of Rochester, Rochester, New York, 14627, USA}
\affiliation{Department of Physics, University of Ottawa, Ottawa ON K1N 6N5, Canada}

\date{\today}

\begin{abstract}

\end{abstract}

\maketitle

\section{Theory of THz single-pixel imaging using electro-optic effect}
The electro-optic (EO) effect can be used to accurately measure the amplitude and phase of THz fields \cite{lee2009principles,zhang2010introduction}. The THz pulses introduce birefringence in EO crystals which changes the polarization of the probe beam. By using the quarter wave plate (QWP) and Wollaston prism, the probe beam will be separated into two beams with equal intensity but orthogonal polarization if there is no overlap with the THz beam. After the THz beam modulates the polarization of the probe beam, the intensity difference $I_s(t)$ between these two separated beams has the following relation with the THz field $E_{THz}(x,y,\omega_{THz},t)$ \cite{lee2009principles}:
\begin{equation}
\begin{split}
  I_s(t) = &\int\int \frac{1}{c} I_0(x,y,\omega_1)\omega_1 L n^3_0(\omega_1) r_{41} |E_{THz}(x,y,\omega_{THz},t)| dxdy,
  \end{split}
\end{equation}
where $c$ is the speed of light in vacuum, $I_0(x,y,\omega_1)$ is the intensity distribution of the probe on the detection crystal. $L$ is the thickness of the detection crystal, and $\omega_1$ is the frequency of probe. $n_0(\omega_1)$ is the refractive index of the detection crystal at optical frequency $\omega_1$, and $r_{41}$ is the EO efficient of the detection crystal. From Eqn. (1), one can see that the THz field can only be detected if it both spatially and temporally coincides with the probe on the detection crystal. Therefore, solely manipulating the probe arm in the spatial domain is equivalent to spatially modulating the THz field in the same manner but leaving the probe field unchanged. Moreover, from the measurement point of view, as long as the polarization of the optical probe is kept, spatially changing the probe field will not change the signal on the balanced detector. Therefore, in the THz computational sampling, it does not matter whether we manipulate the THz field or the probe field. However, even both methods give the same result in principle, encoding patterns on the probe beam can give more benefits as what we discuss in the main paper. Therefore, we can use probe beam encoding to combine computational imaging algorithms and THz imaging techniques.\par

Recently, computational imaging becomes a promising candidate of single-pixel imaging methods. In essence, measuring a $N$ pixel field, i.e. imaging a field, is equivalent to measure an unknown vector $\psi$ in a N dimensional Hilbert space. Therefore, this unknown vector can be decomposed into a set of complete orthogonal basis $H$ in that N dimensional Hilbert space with a coefficient set $\phi$. We can rewrite this decomposition into the matrix form: $\psi = H \phi$, where $\psi$ and $\phi$ are two $N$ dimensional column vectors representing the unknown vector and the coefficients of each basis vector respectively. $H$ is a $N\times N$ matrix and each row represents a basis vector in the Hilbert space. Therefore, one can reconstruct the unknown vector $\psi$ by measuring the coefficients sets $\phi$ in the complete orthogonal basis $H$, and these coefficients can be found as $\phi = H^{-1} \psi$. In our case, this complete orthogonal basis is Hadamard matrix (HM), which can minimize the mean square error of the image \cite{harwit2012hadamard}. HM is a square matrix with $+1$ or $-1$ elements, and the rows of HM are mutually orthogonal. Therefore it satisfies the relation: $H_N H^{T}_N = H^{T}_N H_N = NI_N$, where $H_N$ is the HM in a $N$ dimensional Hilbert space, $H^{T}_N$ is the transpose of $H_N$ and $I_N$ is the $N$ dimensional identity matrix. The HM consists of $+1$ or $-1$ elements, but our SLM can only provide $+1$ or $0$ elements for on and off measurements. Thus, we decompose the HM matrix into $H_{N,1}$ and $H_{N,-1}$ which carries only $+1$ or $0$ elements and $-1$ or $0$ elements respectively: $H_N = H_{N,1}+H_{N,-1} = H_{N,1}- |H_{N,-1}|$. Considering the rules of linear algebra, we have the following relation to reconstruct the image:
\begin{equation}
  \psi = (H_{N,1}- |H_{N,-1}|) \phi,
\end{equation}
where both $H_{N,1}$ and $|H_{N,-1}|$ only carry $+1$ or $0$ elements so that they can be generated from SLM directly. Therefore, in the experiment, we encode the probe beam with $H_{N,1}$ and $|H_{N,-1}|$ patterns sequentially, and record the corresponding signals. To reconstruct the image, one needs to subtract the $|H_{N,-1}|$ patterns weighted by the corresponding signal from $H_{N,1}$ patterns weighted by the corresponding signal. Although this differential method will double the measurement time, it can eliminate the source noise if one immediately shines the $|H_{N,-1}|$ mask right after the corresponding $H_{N,1}$ mask. That is to say, one does not need to introduce any additional power monitor to track the laser power fluctuation. \par
Now we consider an unknown object with intensity distribution $O(x,y)$ placed before the detection crystal. For a given $i_{th}$ HM pattern $I_{i}(x,y,\omega_1)$ which is imaged from SLM plane onto the detection ZnTe crystal with thickness $d = z_1-z_0$, we can find the measured THz total field $P_{s,i}$ from Eqn. (1) as:
\begin{equation}
\begin{split}
    P_{s,i}(t) = \int\int\int_{z_0}^{z_1} \frac{1}{c} I_{i}(x,y,z,\omega_1)\omega_1 L n_0(\omega_1)^3 r_{41}
    |E_{THz}(x,y,z,\omega_{THz},t)| dxdydz,
    \end{split}
\end{equation}
where $E_{THz}(x,y,z,\omega_{THz},t)$ can be found from scalar diffraction theory using angular spectrum:
\begin{equation}
\begin{split}
  E_{THz}(x,y,z,\omega_{THz},t)= \mathfrak{F}^{-1}\{ \mathfrak{F}[O(x,y,0)] exp[i2\pi\sqrt{(k_{THz}/2\pi)^2-f_x^2-f_y^2}z]\}.
  \end{split}
\end{equation}
The $\mathfrak{F}$ and $\mathfrak{F}^{-1}$ represent Fourier Transform and Inverse Fourier Transform respectively. $f_x$ and $f_y$ are spatial frequencies in $x$ and $y$ directions. Here, we have assumed that the THz wave has an uniformly distributed intensity. One can then find the corresponding HM pattern $I_{i}(x,y,z,\omega_1)$ using Eqn. (4). With $I_{i}(x,y,z,\omega_1)$ and $E_{THz}(x,y,z,\omega_{THz},t)$, we can find the signal $P_{s,i}$ for the $i_{th}$ pattern. By repeating this procedure for all patterns, we can get the signal set $P_{s,i}$ which form the coefficient vector $\phi$. This coefficient vector will be used to reconstruct the image vector $\psi$ through Eqn. (2). To show the state of art, the analysis here does not include the broadband nature of THz pulse. However, the simulation we will show already take the spectra contribution into consideration.
\section{Theory of compressed sensing}

\begin{figure}[h!]
\includegraphics[width=1\textwidth,keepaspectratio]{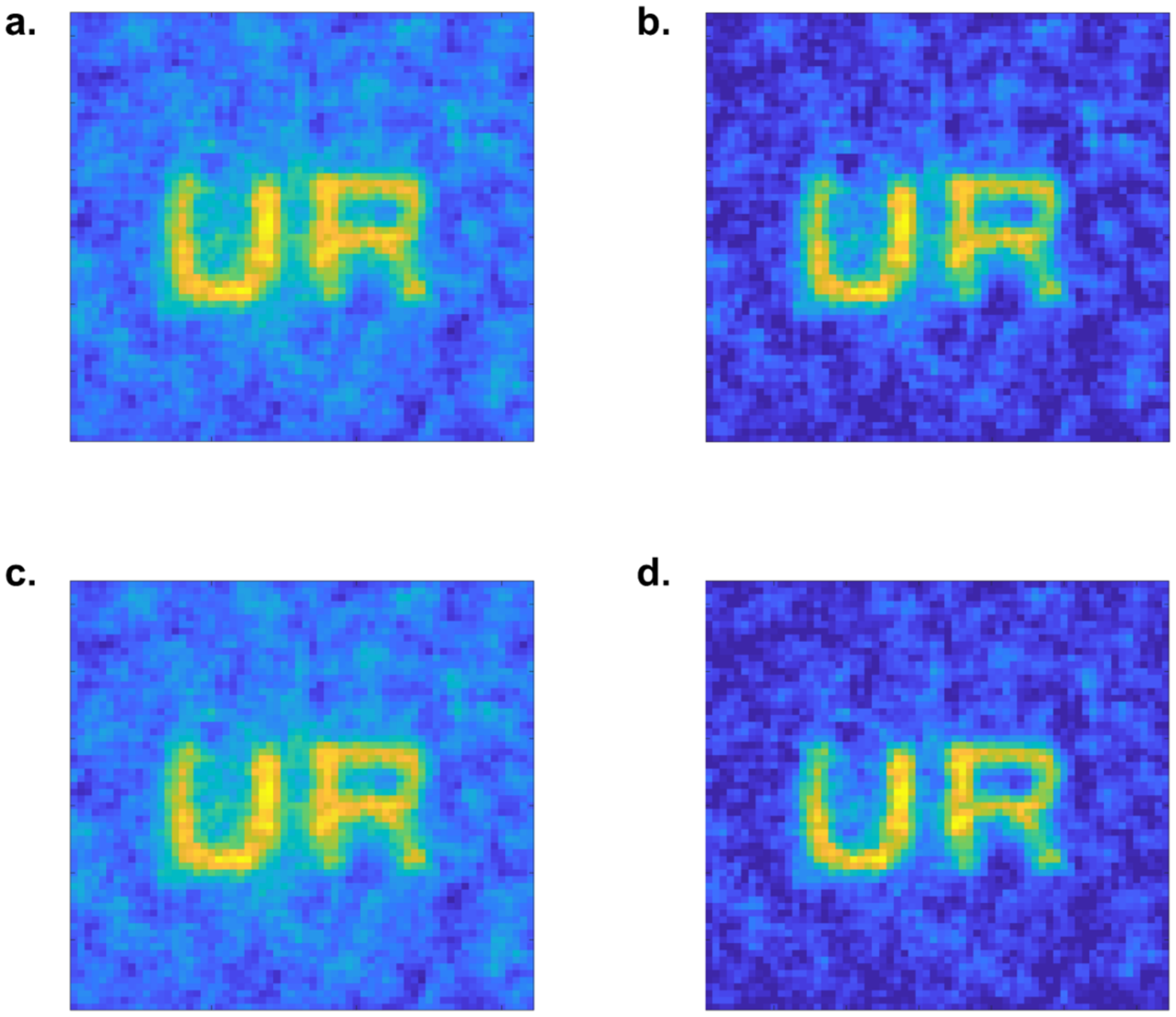}
\caption{Figure (a) to (d) are normalized fields using TV, TV+, TV/L2, and TV/L2+ respectively. The fidelity of TV/L2+ and TV+ is 94.29$\%$ while TV and TV/L2 have fidelities equal to 91.52$\%$.}
\end{figure}
The basic idea of compressed sensing comes from the fact that, for a $N$ dimensional vector $\psi$, it is possible to find an orthogonal basis $T$ that most coefficients in the coefficient set $\phi$ are zero or very small. Due to this sparsity in the vector decomposition, there will no much information loss if we discard these small or zero coefficients. Therefore, one can recover the original vector $\psi$ with a high fidelity using only those large coefficients. Now we assume the THz field can still be represented by a $N$ dimensional column vector $\psi$ in the Hilbert space. In the conventional linear solutions, as what we discuss in the last section, we can write the solution to $\psi$ in the matrix form as: $\phi = Q \psi$, where $Q$ represents the linear transformation matrix representing the basis, and $\phi$ still represents the coefficients set. Now we consider another set of basis represented by $T$, which is assumed to be incoherent with basis $Q$. If we assume that there are only few nonzero coefficients under this transformation $T$, we can recover the $\psi$ using a nonlinear strategy by solving the convex optimization problem with \cite{romberg2008imaging}:
\begin{equation}
  \stackrel[\psi^{'}]{}{\text{min}}\Vert T \psi^{'}\Vert_{\ell_1}, \text{subject to } Q \psi^{'} = \phi,
\end{equation}
where $\Vert  \cdot \Vert_{\ell_1}$ is the 1-norm. If the coherent coefficient between two bases $Q$ and $T$ is very little, i.e. two bases are incoherent, the THz field vector $\psi$ can be reconstructed with $M\geq O[Klog(N)]$ measurements, where $K$ is the number of nonzero components of vector $T \psi$ \cite{candes2007sparsity}. Therefore, we can sub-sample the THz field by the Nyquist-Shannon law but still recover the field with a high fidelity. \par

From the practical point of view, we use the TVAL3 package to solve the minimization problem. The package was provided by Chengbo Li, Wotao Yin and Yin Zhang from Rice University \cite{li2011compressive}. There are 4 different models available and we use TV/L2+ model for our field reconstruction. More than 10 parameters can be adjusted in the model. In Fig. 1, we show the recovered images using different model with same parameter setting. In principle, it is possible to get higher fidelity images by carefully adjusting each parameter of each model. However, that will be trivial and beyond the scope of our current work. From Fig. 1, we can find that 4 different models do not have substantial difference in the image quality. Both TV/L2+ and TV+ have 94.29$\%$ fidelities while TV and TV/L2 have a fidelity equal to 91.52$\%$. \par

\section{Common path interferometer}
If we illuminate an arbitrarily polarized light onto the liquid crystal SLM, only the horizontally polarized portion gets modulated. Therefore, through the utilization of this property, we will use common path interferometer to transfer the phase-only spatial patterns into intensity-only patterns. In our case, the axis of HWP 1 in Fig. 1(a) is fixed at $22.5^{\circ}$ so that the polarization of the optical probe beam is changed from horizontal polarization to diagonal polarization ($45^{\circ}$ to the horizontal direction). We can write the polarization state $|D\rangle$ before the SLM as:
\begin{equation}
  |D\rangle = \frac{1}{\sqrt{2}}(|H\rangle+|V\rangle),
\end{equation}
where $|H\rangle$ and $|V\rangle$ represent the horizontal and vertical polarization states respectively. The phase masks on the SLM are binary masks with $0$ and $\pi$ phase delay. Therefore, after reflected by the SLM, the polarization of areas with $\pi$ phase delay is in anti-diagonal polarization state $|A\rangle$. If we assume the binary spatial phase mask on the SLM can be represented by term $\psi (x,y)$, we will find the state after the SLM $|D\rangle_{SLM}$ in the form:
\begin{equation}
  |D\rangle_{SLM} = \frac{1}{\sqrt{2}}(|H\rangle+exp(i\psi (x,y))|V\rangle).
\end{equation}
Right after the SLM, we use a polarizer with its axis at $45^{\circ}$ to interfere the $|H\rangle$ and $|V\rangle$ components:
\begin{equation}
  \langle D|D\rangle_{SLM} = \frac{1}{2}(1+exp(i\psi (x,y))).
\end{equation}
For those parts with the phase delay equal to $\pi$, Eqn. (4) will become 0, and the $0$ phase delay parts will become unity. Hence, we get a binary intensity pattern from the phase-only SLM using common path interferometer. \par
Even though the intensity-only device such as digital micromirror device (DMD) has a much higher speed (kHz level), there are two advantages of using the phase-only SLM. Since the micromirror array, which basically works like a grating, will lead to strong diffraction to the incoming beam, these DMDs usually have low efficiencies. The other drawback is that, due to the reflection nature of micromirror array, it will spatially shear the pulse front. In our test, the duration of a pulse can be stretched from 200 $fs$ to 4 $ps$. Therefore, in order to use the DMD, one has to compensate this spatial shear effect. A recently published work also discussing this effect and applying it to THz pulse generation when we prepare the draft \cite{murate2018adaptive}.

\section{Resolution estimation}
Here we show some numerical simulation results to estimate the resolution limit of this technique as by varying different parameters. The simulation is based on the theory shown in the first section. As mentioned in the main paper, we investigate the influences of three different parameters: the thickness of detection crystal, the pixel size and the central wavelength of the THz pulse.\par

\begin{figure}[h!]
\includegraphics[width=1\textwidth,keepaspectratio]{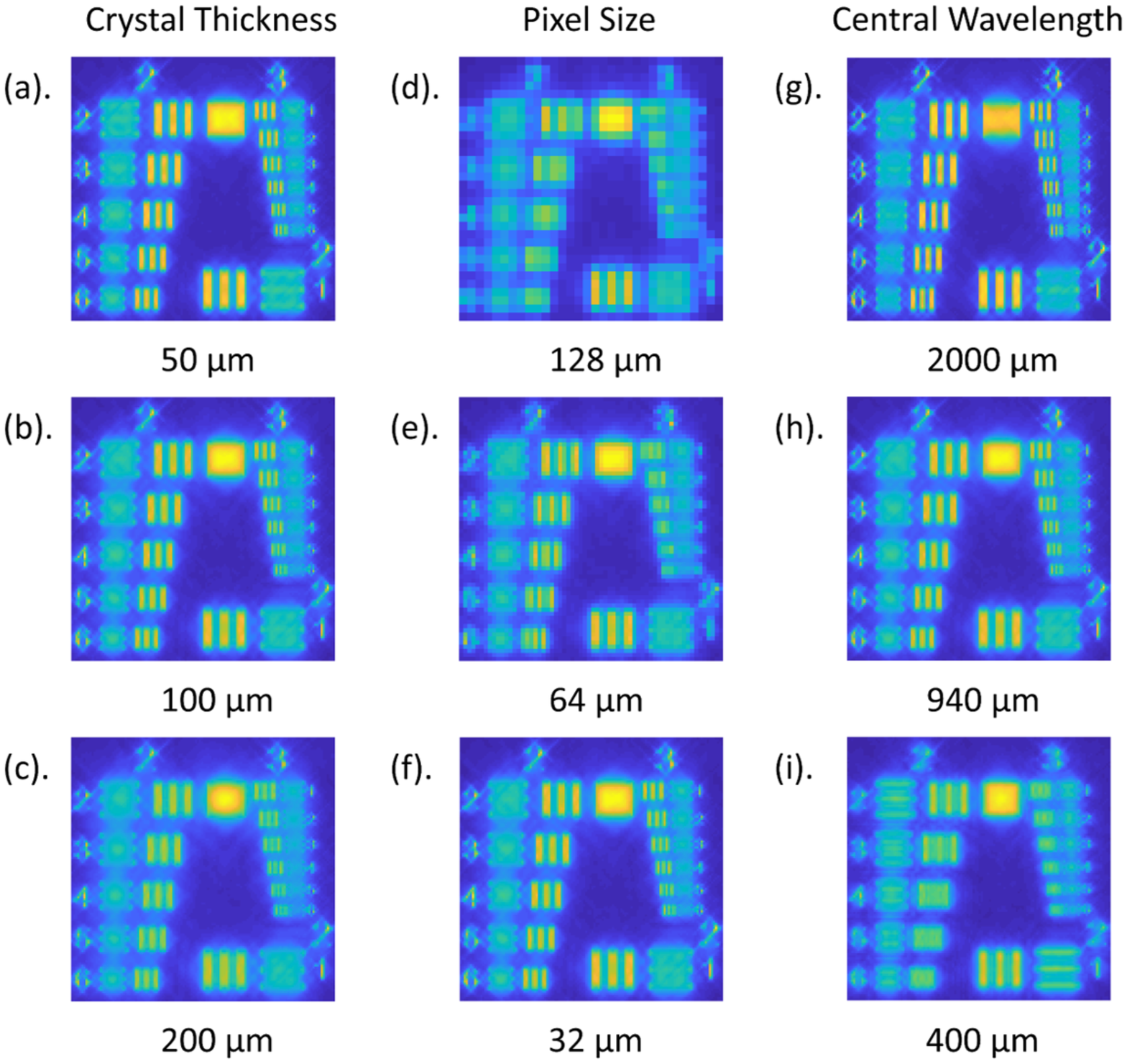}
\caption{Figure (a) to (c) are normalized results with different crystal thicknesses. Figure (d) to (f) are normalized results with different pixel sizes. Figures (g) to (i) are normalized results with different central wavelengths.}
\end{figure}

Three different thicknesses of detection ZnTe crystal are chosen with a 32 $\mu m$ pixel size and 128 pixels: 50 $\mu m$, 100 $\mu m$ and 200 $\mu m$. The central wavelength is 940 $\mu m$. As shown in Fig. 2 (a)-(c), the reconstructed field with 50 $\mu m$ crystal thickness is the most clear one. By calculating the contrast of each element using the same method as mentioned in the main draft, we find that with 50 $\mu m$ and 100 $\mu m$ detection crystal thicknesses, even the element 3-6 (d = 35 $\mu m$) can be fully resolved with a 35.36$\%$ contrast and a 28.47$\%$ contrast respectively. As a comparison, the resolution in Fig. 2 (c) is only 88 $\mu m$ with a 20.79$\%$ contrast. Therefore, a thin detection crystal can significantly increase the resolution. However, from the Eqn. (3), a thin detection crystal will lead to a smaller signal. Hence, for applications requiring high SNRs, one needs to carefully choose the thickness of the detection crystal to balance the SNR and resolution.\par

Three different pixel sizes are simulated under a 100 $\mu m$ thick detection crystal with a 940 $\mu m$ central wavelength THz pulse are simulated: 128 $\mu m$, 64 $\mu m$ and 32 $\mu m$. Since our field size is fixed, the number of pixels in each case then becomes 32, 64 and 128 respectively. The recovered figures are shown in Fig. 2(d)-(f) respectively. The resolution in Fig. 2(d) is 125 $\mu m$ and the contrast is only 22.08$\%$. As the comparison, the resolution in Fig. 2(e) is 88 $\mu m$ with a 25.96$\%$ contrast while the resolution in Fig. 2(f) is less than 35 $\mu m$ as we discussed in previous paragraph. These results shows that the pixelization effect can significantly reduce the resolution and contrast, and one can expect a better resolution if we use a smaller pixel size. Therefore, optical SLMs are more favorable than THz SLMs in computational imaging since the pixel size is much smaller, which indicates that our method can provide better performance than those methods with THz SLMs. \par

Three different central wavelength with a same spectrum shape are analyzed with a 100 $\mu m$ thick detection crystal: 2000 $\mu m$, 940 $\mu m$, 400 $\mu m$. The number of pixels is 128 so that the pixel size is 32 $\mu m$. Note that even the central wavelength has shifted, the shapes of the spectrum are same for all three cases. Intuitively speaking, Fig. 2(g) (2000 $\mu m$ central wavelength) has the best contrast while Fig. 2(i) is the worst. By calculating the contrast of the elements in all three figures, we find that the resolution in Fig. 2(i) is only 125 $\mu m$ with a 21.05$\%$. That is to say, in the sub-wavelength region, the resolution in Fig. 2(i) is worse than the resolution in Fig. 2(c) even the detection crystal is much thinner. However, the resolution in Fig. 2(g) should be much less than 35 $\mu m$ since the contrast of element 3-6 is found to be 48.11$\%$, which is much higher than the contrast of the same element in Fig. 2(a). Therefore, in the near-field region, a longer central wavelength pulse will provide a better resolution even when the detection crystal is thicker. The conclusion here is very different from the conclusion in far-field imaging where a shorter wavelength is always desired for a higher resolution.

\section{Improved resolution compared to near-field EO imaging}
\begin{figure}[h!]
\includegraphics[width=1\textwidth,keepaspectratio]{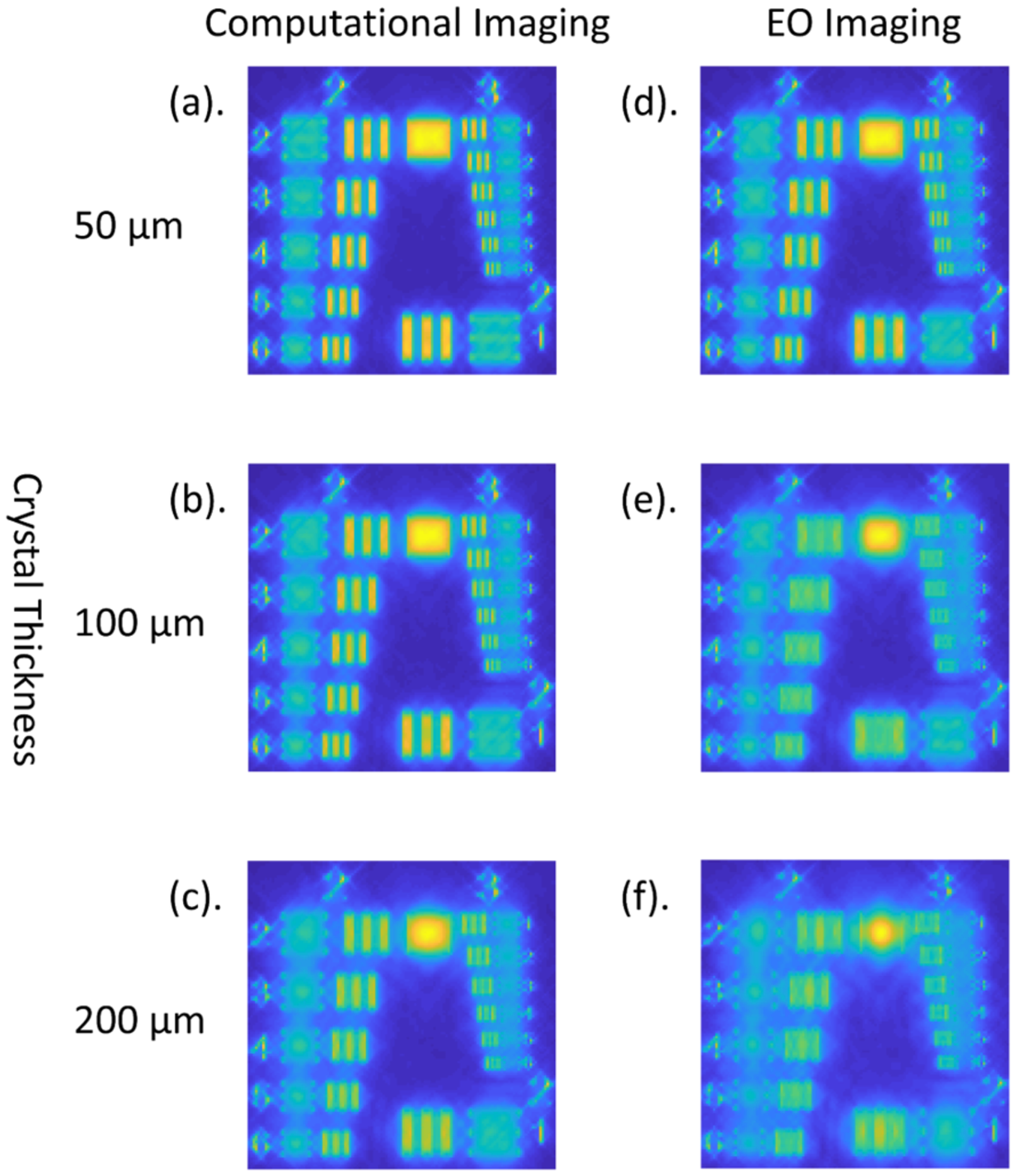}
\caption{Figure (a) to (c) are normalized results using computational imaging methods. Figure (d) to (f) are normalized results using EO imaging methods. The figure (a) and (d) both have a 50 $\mu m $ thick detection crystal while the crystal thicknesses in figure (b) and (e) are both 100 $\mu m$. The crystal thicknesses in figure (c) and (f) are 200 $\mu m$.}
\end{figure}
The conventional near-field EO imaging utilizes the similar detection method as ours \cite{chen2000electro,wang2010terahertz}. One can use an optical CCD array to retrieve real time images of the THz field. Unlike other near-field imaging techniques, this EO imaging technique also provides noninvasive measurements with a concise and reliable setup. However, the resolution of this technique is not as good as our approach especially when the crystal goes thicker. The resolution limitations in both cases highly depend on the thickness of the detection crystal. However, this factor has less impact in our case because we measure the total electrical field of each spatial pattern as shown in Eqn. (3), which is an accumulation result through the whole thickness of detection crystal. Therefore, we can find a position $z'$ where the product of THz field, the $i_{th}$ spatial pattern and the thickness of the crystal $d$ is exactly equivalent to the integral in Eqn. (3). That is to say, we can rewrite the Eqn. (3) into another form:
\begin{equation}
  \begin{split}
 I_{s,i}(t) = &\int\int\int_{z_0}^{z_1} \frac{1}{c} I_{i}(x,y,z,\omega_1)\omega_1 L n_0^3(\omega_1) r_{41} \times \\
    &|E_{THz}(x,y,z,\omega_{THz},t)| dxdydz \\
    = &\int\int \frac{1}{c} z' I_{i}(x,y,z',\omega_1)\omega_1 L n_0^3(\omega_1) r_{41} \times \\
    &|E_{THz}(x,y,z',\omega_{THz},t)| dxdy,
    \end{split}
    \end{equation}
where $E_{THz}(x,y,z',\omega_{THz})$ and $I_{i}(x,y,z',\omega_1)$ are THz field and $i_{th}$ spatial pattern at distance $z'$ respectively. Note that $z'$ is less than $z_1$ so that the THz field at $z'$ is less diffracted than the field at $z_1$. Therefore, what the computational imaging recovers is not the field at the end of detection crystal but the field at position $z'$. However, in the EO near-field imaging case, what the camera measures is the transverse structure of the THz field at the rear surface of detection crystal (i.e. $z_1$), which is more blurred due to the relatively stronger diffraction. This difference in the measurement favors the superior resolution in our sampling technique by sacrificing the image acquisition time. Another drawback in the near-field EO imaging is the requirement of high power lasers \cite{mittleman2018twenty}. As what we discuss in the last section, a thinner crystal can give better resolution but worse contrast and SNR. Therefore, to get high resolution images with a good contrast, a high energy laser is used to provide a strong THz field. However, in our case, high contrast images can be retrieved with a low intensity pump laser and a extremely weak probe.\par
To intuitively show the comparison, we compare the reconstructed images from computational imaging and the images from near-field EO imaging with different thicknesses of the crystal. As shown in Fig. 3(a)-(c), three different images with thicknesses 50 $\mu m$, 100 $\mu m$ and 200 $\mu m$ are reconstructed. To keep the comparison fair, this set of images all have 128 pixels with 32 $\mu m$ pixel size. After calculating the contrast of each element, the element 3-6 (d = 35 $\mu m$) can still be fully resolved in Fig. 3(d) but with a 26.11$\% $ contrast, which is much lower than the contrast using computational imaging method in Fig.3(a). For Fig. 3(e) and (f), there is no element can be resolved. That is to say, our computational imaging method can significantly improve the resolution especially when the detection crystal is quite thick, which matches our prediction. Therefore, for applications requiring a thick detection crystal, our approach can provide much better performance than the conventional EO imaging. Therefore, under the same conditions, our computational method is superior than the conventional EO imaging in resolution and contrast with a sacrifice of imaging speed.


%